\documentclass[12pt,a4paper]{article}
\usepackage{amsmath}
\usepackage{graphics}

\newcommand{\hepth}[1]{{\tt hep-th/#1}}
\newcommand{\plb}[3]{Phys.Lett. {\bf B#1} (#2) #3}

\newcommand{\ep}{\varepsilon} 
\newcommand{\half}{\frac{1}{2}}
\newcommand{\K}{$\tilde{\cal K}$}

\newcommand{\nn}{\nonumber}

\newcommand{\p}{\vspace{6pt}\noindent}
\advance\textheight by \topskip
\makeatletter

\@addtoreset{equation}{section}
\def\section{\@startsection {section}{1}{\z@}{-8.5ex plus -1ex minus
 -.2ex}{3.3ex plus .2ex}{\large\bf}}
\def\subsection{\@startsection{subsection}{2}{\z@}{-3.25ex plus
 -1ex minus -.2ex}{1.5ex plus .2ex}{\bf}}
\def\subsubsection{\@startsection{subsubsection}{3}{\z@}{-3.25ex plus%
 -1ex minus -.2ex}{1.5ex plus .2ex}{\sl}}

\begin{document}
\begin{titlepage}
\vspace*{-2cm}
\begin{flushright}

 hep-th/0401020 \\

\end{flushright}

\vspace{0.3cm}

\begin{center}
{\Large {\bf Affine Toda field theories with defects }}\\
\vspace{1cm} {\large P.\ Bowcock\footnote {\noindent E-mail: {\tt
peter.bowcock@durham.ac.uk}}, E.\ Corrigan\footnote{\noindent
E-mail: {\tt ec9@york.ac.uk}} and
C.\ Zambon\footnote{\noindent E-mail: {\tt cz106@york.ac.uk}} }\\
\vspace{0.3cm} {${}^a$}\ \em Department of Mathematical
Sciences\\ University of Durham\\ Durham DH1 3LE, U.K.\\
\vspace{0.3cm} {${}^{b,c}\ $\em\it Department of
Mathematics \\ University of York\\
York YO10 5DD, U.K. }\\ \vspace{1cm} {\bf{ABSTRACT}}
\end{center}

\begin{quote}
A Lagrangian approach is proposed and developed to study defects within affine Toda
field theories. In particular, a suitable
Lax pair is constructed together with examples of conserved charges.
It is found that only
those models based on  $a_r^{(1)}$ data appear to allow defects preserving
integrability. Surprisingly, despite the explicit breaking of
Lorentz and translation invariance,
modified forms of both energy and momentum are conserved.
Some, but apparently not all, of the higher spin conserved charges  are also
preserved after the addition of  contributions
from the defect. This fact is illustrated by noting how defects may preserve
a modified form of just one of the spin 2 or spin -2 charges but not
both of them.
\end{quote}

\vfill
\end{titlepage}

\section{Introduction}
\label{s:intro} In a recent article, based principally on the
examples of the sinh-Gordon, Liouville, or free scalar field
models \cite{bczlandau}, it was pointed out that field theories in
$1+1$ dimensions may have internal boundary conditions with
interesting consequences. Typically, at an internal boundary the
classical field will have a discontinuity, hence the name
`defect', yet energy and momentum will be conserved once they have
been suitably modified. In addition, any number of defects may be
placed along the $x$-axis, and each will introduce additional
parameters to the model. The purpose of this article is to explore
the consequences of integrability in the presence of defects.

\p
The quantum version of a situation similar to this has been examined before and
imposing the requirements of integrability was found to be highly
restrictive. The investigation was pioneered by Delfino,
Mussardo and Simonetti some years ago \cite{Delf94} and there has
been renewed interest in it recently
\cite{Castro-Alvaredo2002, Mintchev02}.

\p
Over some years, there has been interest in the study of integrable
classical or quantum field theories restricted to a half-line, or
interval, by imposing integrable boundary conditions, say at $x=0$,
see for example
\cite{Fring93,Ghosh94a, Ghosh94b, Cor94a, Cor95, MacI95, Bow95}.
The simplest
situation, which is also the best
understood, contains a real self-interacting scalar field $\phi$ with
either a periodic ($\cos$), or non-periodic ($\cosh$) potential.
The sinh-Gordon model can be restricted to the left
half-line $-\infty\leq x\leq 0$, without losing integrability, by
imposing the boundary condition
\begin{equation}\label{bc}
 \left. \partial_x\phi\right|_{x=0}=\frac{\sqrt{2}m}{\beta}
 \left(\ep_0e^{-\frac{\beta}{\sqrt{2}}\phi(0,t)}-
\ep_1e^{\frac{\beta}{\sqrt{2}}\phi(0,t)}\right) ,
\end{equation}
where $m$ and $\beta$ are the bulk mass scale and coupling constant,
respectively, and $\ep_0$ and $\ep_1$ are two additional parameters
\cite{Ghosh94a, MacI95}.  This set of boundary conditions
generally breaks the reflection symmetry $\phi\rightarrow -\phi$
of the  model although the symmetry is explicitly preserved when
$\ep_0=\ep_1\equiv\ep$. The restriction of the sinh-Gordon model
to a half-line is a considerable complication, and renders the
model more interesting than it appears to be in the bulk. This is
because generally there will be additional states in the spectrum
associated with the boundary, together with a set of reflection
factors which must be compatible with the bulk S-matrix
(see \cite{ Ghosh94a, Ghosh94b, Cor95, Zam79}). The
weak-strong coupling duality enjoyed by the bulk theory emerges in
a new light \cite{Cor97, Chen00, Cor00}.

\p
Integrability in the
bulk sinh-Gordon model requires the existence of conserved quantities
labelled by odd spins $s=\pm 1, \pm 3, \dots$, and some of these should
survive even in the presence of boundary conditions. Since boundary
conditions typically violate translation invariance, it would be expected that
the `momentum-like' combinations of conserved quantities should not be
preserved. However, the `energy-like' combinations, or some subset of them,
might remain conserved, at least when suitably modified (see \cite{Ghosh94a}
for the paradigm). As is the case for the theory restricted to a half-line,
it was reported in \cite{bczlandau} that
the spin three charge already supplies the most general restrictions on
the internal boundary condition in the presence of a defect.
The Lax pair approach developed in
\cite{Bow95}, adapted to the new context, was used to re-derive the
boundary conditions, thereby demonstrating that the preservation of
higher spin energy-like charges imposed no further restrictions on the
internal boundary conditions. More surprisingly, it was noted that
after suitable modification momentum was conserved.

\p In this article, it is intended to give a fuller discussion of defects
from a Lagrangian point of view, and to examine the possibilities
within the context of affine Toda field theories. The basic ideas will
be illustrated by a free massive real scalar field since for that case
there are no complications arising from the requirements of integrability.
Later, the Lax pair will be used to establish the existence of defects
within a certain class of affine Toda field theory - those based on the
root data of $a_r^{(1)}$ - the simplest example of these being the
already-discussed sinh-Gordon model, based on $a_1^{(1)}$ data.

\section{Real scalar field}
\label{s:fsf}

\p For convenience, the field in the region $x>0$ will be denoted $\psi$
and the field in the region $x<0$ will be denoted $\phi$; in principle,
each could have its own bulk potential denoted $W(\psi)$ and $V(\phi)$,
respectively. The portion of the Lagrangian associated with
the defect is taken to be
\begin{equation}\label{freedefect}
{\cal L}_D=\delta(x)\left[\frac{1}{2}\left(\phi\partial_t\psi -
\psi\partial_t\phi\right)-{\cal B}(\phi,\psi)\right],
\end{equation}
where ${\cal B}(\phi,\psi)$ is the defect potential, and therefore the defect
conditions at $x=0$ are:
\begin{eqnarray}\label{freeconditions}
 \nonumber \partial_x\phi &=& \partial_t\psi - \frac{\partial{\cal B}}{\partial\phi}\\
  \partial_x\psi &=& \partial_t\phi + \frac{\partial{\cal B}}{\partial\psi}.
\end{eqnarray}
Consider the time derivative of the standard bulk momentum, which is not
expected to be preserved because of the lack of translation invariance.
Then, ignoring contributions from $\pm\infty$, this may be written in terms
of the fields and their derivatives evaluated at the boundary $x=0$:
\begin{eqnarray}\label{Pdot}
  \nonumber \frac{dP}{dt} &=& \frac{d}{dt}\left(\int_{-\infty}^0 dx\,
  (\partial_t\phi\partial_x\phi) + \int_0^\infty dx\,
  (\partial_t\psi\partial_x\psi)\right)\\
  \phantom{\frac{dP}{dt}} &=&\frac{1}{2} \left[(\partial_x\phi)^2 +(\partial_t\phi)^2
  - (\partial_x\psi)^2 -(\partial_t\psi)^2 -2V(\phi) +2 W(\psi)\right]_{x=0}.
\end{eqnarray}
Using the boundary conditions \eqref{freeconditions}, the combination
of terms evaluated at the boundary may be rewritten
\begin{equation}
 \nonumber    -\partial_t\psi\frac{\partial{\cal B}}{\partial\phi}-
 \partial_t\phi\frac{\partial{\cal B}}{\partial\psi}-V(\phi)+W(\psi)
 +\half\left(\frac{\partial{\cal B}}{\partial\phi}\right)^2-
 \half\left(\frac{\partial{\cal B}}{\partial\psi}\right)^2,
 \end{equation}
 which in turn is a total time-derivative of a functional of $\phi(0,t)$ and
 $\psi(0,t)$, provided
 \begin{eqnarray}\label{Bconditions}
   \frac{\partial^2{\cal B}}{\partial \phi^2} - \frac{\partial^2{\cal B}}
   {\partial \psi^2}&=& 0, \\
   \half\left(\frac{\partial{\cal B}}{\partial\phi}\right)^2-
 \half\left(\frac{\partial{\cal B}}{\partial\psi}\right)^2 &=& V(\phi)-W(\psi).
 \end{eqnarray}
 There are many solution to equations \eqref{Bconditions} but, for free fields,
 the simplest would be to take
 \begin{eqnarray}\label{simplestcase}
   V(\phi)&=&\half m^2\phi^2,\qquad  W(\psi)=\half m^2\psi^2 \\
   {\cal B} &=&\left[ \frac{m\lambda}{4}(\phi+\psi)^2+\frac{m}{4\lambda}
   (\phi-\psi)^2\right]_{x=0},
 \end{eqnarray}
 where $\lambda$ is a free parameter. Clearly, there is no requirement
 for $\phi$ and $\psi$ to match at $x=0$,
 except in the limit $\lambda\rightarrow 0$.
 With the expressions \eqref{simplestcase} the combination
\begin{equation}\label{totalmomentum}
   {\cal P}= P+\left[ \frac{m\lambda}{4}(\phi+\psi)^2-\frac{m}{4\lambda}
   (\phi-\psi)^2\right]_{x=0}
\end{equation}
is readily checked to be a conserved quantity, as is the total energy
\begin{equation}\label{totalenergy}
    {\cal E}= E+\left[ \frac{m\lambda}{4}(\phi+\psi)^2+\frac{m}{4\lambda}
   (\phi-\psi)^2\right]_{x=0}.
\end{equation}
Indeed, in light-cone coordinates:
\begin{equation}\label{EplusorminusP}
    {\cal P}_{\pm 1}={\cal E}\pm {\cal P}= E\pm P+\left[ \frac{m\lambda^{\pm 1}}{2}
    (\phi\pm\psi)^2\right]_{x=0},
\end{equation}
indicating that the parameter $\lambda$ introduced at the defect,
although breaking Lorentz invariance, has naturally the character
of a spin one quantity.

\p
For the linear system, the defect simply causes a delay. Thus with,
\begin{equation}\label{}
\nn \phi=e^{-i\omega t+ikx}+R e^{-i\omega t -ikx},\ x<0; \quad
    \psi=Te^{-i\omega t+ikx},\ x>0,
\end{equation}
the boundary conditions require
\begin{equation}\label{}
\nn R=0,\ T=-\, \frac{ik+\frac{m}{2}\left(\lambda+\frac{1}{\lambda}\right)}
{i\omega+\frac{m}{2}
\left(\lambda-\frac{1}{\lambda}\right)}.
\end{equation}
Clearly, neither the fields nor their spatial derivatives are continuous
through the defect.

 \p It is interesting to note
 (see \cite{bczlandau})
 that if the conditions \eqref{freeconditions} were regarded as a pair of
 differential equations in the bulk then the additional conditions on the
 boundary potential ${\cal B}$, summarised in \eqref{Bconditions}, would
 guarantee the pair to be a B\"acklund transformation leading to
 \begin{equation}\label{bulkequations}
    \partial^2\phi = -\frac{\partial V}{\partial \phi}, \qquad
    \partial^2\psi = -\frac{\partial W}{\partial \psi}.
\end{equation}
Following from this remark, it is to be expected that any pair of fields
$\phi,\ \psi$ related by a B\"acklund transformation \cite{Back1882} should be able to
support a defect. In \cite{bczlandau} it was pointed out that besides
free massive fields this is indeed
the case for sinh-Gordon, Liouville and massless free fields, assembled
in suitable combinations. The close association of the B\"acklund transformation
with a defect offers new insight into the B\"acklund transformation itself.

\p In this article the analysis is extended to other affine Toda
field theories.

\section{Lax pair for affine Toda field theory with a defect}
\label{s:laxpair}

\p The starting point for this discussion will be the standard affine
Toda field theory lagrangian (see \cite{Cor94nd})
together with a defect contribution, which
in the first instance is located at $x=0$. The notation used above for
the fields in the two regions $x<0$ and $x>0$ will be maintained although
in the case of affine Toda field theory the fields will be multi-component.
Bearing this in mind, a fairly general Lagrangian, including
a defect contribution, would be
\begin{equation}\label{affinedefect}
    {\cal L}=\theta(-x){\cal L}_\phi +\theta(x){\cal L}_\psi
    +\delta(x)\left(\half\phi E\partial_t\phi +\phi D\partial_t\psi
    +\half\psi F\partial_t\psi - {\cal B}(\phi,\psi)\right),
\end{equation}
where $D,\ E$ and $F$ are matrices independent of $\phi$ and $\psi$.
Omitting total time derivatives requires $E$ and $F$ to be antisymmetric.
More general possibilities might allow these to be functions of the
fields, and might also allow for other than linear terms in time derivatives.
For models on a half-line, the latter possibility has been
considered by Baseilhac and Delius \cite{Bas01}.

\p
Following from the Lagrangian \eqref{affinedefect}, the defect
conditions at $x=0$ are:
\begin{eqnarray}\label{affineconditions}
 \nonumber \partial_x\phi - E\phantom{^T}\partial_t\phi -D\, \partial_t\psi +
  \nabla_\phi{\cal B} &=& 0 \\
  \partial_x\psi - D^T\partial_t\phi +F\,  \partial_t\psi -
  \nabla_\psi{\cal B} &=& 0.
\end{eqnarray}

\p Using the ideas presented in \cite{Bow95} it is straightforward to
set up a Lax pair which will automatically incorporate the boundary conditions
\eqref{affineconditions}. In the bulk, the Lax pair for affine Toda field theory
is well-known and has the following form:
\begin{eqnarray}\label{laxpairsb}
a_t&=& \frac{1}{2}\left[\partial_x\phi\cdot {\bf H}+
\sum_i\sqrt{m_i}e^{\alpha_i\cdot\phi/{2}}\left(\lambda E_{\alpha_i}-\frac{1}{\lambda}
E_{-\alpha_i}\right)\right] \nonumber\\
a_x&=&\frac{1}{2}\left[\partial_t\phi\cdot {\bf H}+
\sum_i\sqrt{m_i}e^{\alpha_i\cdot\phi/{2}}\left(\lambda E_{\alpha_i}+\frac{1}{\lambda}
E_{-\alpha_i}\right)\right],
\end{eqnarray}
 where ${\bf H}$ are the generators in the Cartan subalgebra of the
semi-simple Lie algebra whose simple roots are $\alpha_i,\ i=1,\dots, r$, and
$E_{\pm\alpha_i}$ are the generators corresponding to the simple roots or
their negatives. The additional root, given by $\alpha_0=
-\sum_i n_i\alpha_i$, is the (Euclidean part of) the additional root
in the Dynkin-Ka\v c diagram and is appended to the set of simple roots.
The two expressions \eqref{laxpairsb} are readily checked to
be a Lax pair  using additional facts about the Lie algebra commutation
relations (for more details about this, and further references, see
\cite{Olive85}):
\begin{equation}\label{Liealgebrarelations}
    [{\bf H}, E_{\pm\alpha_i}]=\pm\alpha_i E_{\pm\alpha_i}, \quad [E_{\alpha_i},
    E_{-\alpha_j}]=\delta_{ij}\frac{2\alpha_i}{\alpha_i^2}\cdot{\bf H},
\end{equation}
and $ m_i={n_i\alpha_i^2}/2$. That is
\begin{equation}\label{}
    \partial_t a_x - \partial_x a_t +[a_t,a_x]=0\quad\Leftrightarrow\quad
    \partial^2 \phi = -\nabla_\phi V(\phi),\quad V(\phi)=\sum_{i=0}^r
    n_ie^{\alpha_i\cdot\phi}.
\end{equation}
For ease of notation, the coupling $\beta$ and the bulk mass
parameter $m$ are omitted but may be reinstated by appropriate
rescalings. As usual, there is a family of Lax pairs labelled by
the spectral parameter $\lambda$.

\p
To incorporate a defect at $x=0$, the Lax pair may be adapted as follows
\cite{Bow95}. Extend the region $x<0$ to $x<b,\ b>0$ and the region $x>0$
to $x>a$, $a<0$ (in other words, consider two overlapping regions $R^<$
and $R^>$ each
containing the defect), and in each region define a new Lax pair:
\begin{eqnarray}\label{modifiedlax}
 \nonumber R^<:\qquad \widehat a_t^{\, <} &=& a_t(\phi) -\half\theta (x-a)\left(
          \partial_x\phi - E\phantom{^T}\partial_t\phi -D\, \partial_t\psi +
  \nabla_\phi{\cal B}\right)\cdot {\bf H} \\
\nonumber  \phantom{R^<}\quad \widehat a_x^{\, <} &=& \theta(a-x)a_x(\phi) \\
\nonumber   R^>:\qquad \widehat a_t^{\, >} &=& a_t(\psi) -\half\theta (b-x)\left(
          \partial_x\psi - D^T\partial_t\phi +F\, \partial_t\psi -
  \nabla_\psi{\cal B}\right)\cdot{\bf H} \\
  \phantom{R^>}\quad \widehat a_x^{\, >} &=& \theta(x-b)a_x(\psi).
\end{eqnarray}
Checking directly reveals that eqs\eqref{modifiedlax} yield both the
equations of motion and defect conditions (displaced to $x=a$ and $x=b$, respectively)
for the fields in the two regions $x<a$ and $x>b$. In the overlap region
$a<x<b$, the components $\widehat a_x^{\, >},\ \widehat a_x^{\, <}$ vanish, implying
the other components, $\widehat a_t^{\, >},\ \widehat a_t^{\, <}$, must be
$x$-independent,
in turn implying that both $\phi$ and $\psi$ are independent of $x$
throughout the overlap. On the other hand, maintaining the zero-curvature
condition within the overlap also requires the two components
$\widehat a_t^{\, >}$ and $\widehat a_t^{\, <}$ to be
related by a gauge transformation:
\begin{equation}\label{overlapgauge}
    \partial_t {\cal K}={\cal K}a_t^{\, >}-a_t^{\, <}{\cal K},
\end{equation}
where ${\cal K}$ is a matrix of dimension equal to the dimension of the
representation chosen for the Lie algebra generators
${\bf H},\ E_{\pm\alpha_i}$. With the given boundary conditions, ${\cal K}$
will be $t$-dependent so it is convenient to make the following change:
\begin{equation}\label{changeofK}
    {\cal K}=e^{-\half{\bf H}\cdot(E\phi +D\psi)}\, \tilde {\cal K}\,
    e^{\half{\bf H}\cdot(D^T\phi -F\psi)}.
\end{equation}
\p
Assuming $\tilde{\cal K}$ is $t$-independent and using the explicit
expressions for the Lax pair, eq\eqref{overlapgauge} provides a set
of equations constraining the defect potential ${\cal B}$ together with the
group element $\tilde{\cal K}$:
\begin{eqnarray}\label{Kequation}
\nonumber \tilde{\cal K}\, {\bf H}\cdot\nabla_\psi{\cal B}+
{\bf H}\, \tilde{\cal K}\cdot\nabla_\phi{\cal B} &=& \sum_{i=0}^r\sqrt{m_i}
\left[\lambda \left(E_{\alpha_i}\tilde{\cal K}e^{\half\alpha_i\cdot(\phi +E\phi
+D\psi)}\right.\right.\\
\nonumber &&\phantom{2\sum_{i=0}^r\sqrt{m_i}
(\lambda}-\left.\tilde{\cal K}E_{\alpha_i}e^{\half\alpha_i\cdot(\psi-F\psi+
D^T\phi)}\right)\\
\nonumber &&\phantom{\sum_{i=0}^r\sqrt{m_i}(}-\frac{1}{\lambda}
\left(E_{-\alpha_i}\tilde{\cal K}
e^{\half\alpha_i\cdot(\phi-E\phi-D\psi)}\right.\\
&&\phantom{2\sum_{i=0}^r\sqrt{m_i}(}-\left.\left.\tilde{\cal K}E_{-\alpha_i}
e^{\half\alpha_i\cdot(\psi+F\psi -D^T\phi)}\right)\right].
\end{eqnarray}
Bearing in mind  the defect potential ${\cal B}$ does not depend
on the spectral parameter, and that $\tilde{\cal K}$ is not
expected to depend on the fields in either region (since it could
not be $t$-independent otherwise), it turns out
eq\eqref{Kequation} is extremely strong and severely limits the
possible defects. Note, $\tilde{\cal K}$ has a certain obvious
arbitrariness since it may be multiplied by any function of
$\lambda$, commuting with all generators of the Lie algebra,
without affecting \eqref{Kequation}.

\p
For example, if it is supposed that \K\ has a finite limit as
$\lambda\rightarrow \infty$, with \K$(\infty)=1$, then the terms of order $\lambda$
on the right hand side of \eqref{Kequation} must cancel. For that to be the case,
given the fields and root data are real, the exponential functions
appearing as coefficients of the Lie algebra generators must match exactly
for each $i$. In other words,
\begin{equation}\label{matching}
    \alpha_i\cdot(1+E-D^T)\phi = \alpha_i\cdot(1-F-D)\psi, \quad i=0,\dots,r.
\end{equation}
If it is further supposed that the defect is `maximal', in the sense that
the magnitude of the defect is not prescribed  and there
is no relation between the field values $\phi(t,0)$ and $\psi(t,0)$, then
since the simple roots are linearly independent the three matrices $D,E$ and $F$
must satisfy
\begin{equation}\label{matrixrelations}
    E=F=1-D,\quad D+D^T=2.
\end{equation}
Incorporating these relations into \eqref{Kequation} gives
\begin{eqnarray}\label{newKequation}
\nonumber \tilde{\cal K}\, {\bf H}\cdot\nabla_\psi{\cal B}+
{\bf H}\, \tilde{\cal K}\cdot\nabla_\phi{\cal B} &=& \sum_{i=0}^r\sqrt{m_i}
\left[\lambda \left(E_{\alpha_i}\tilde{\cal K}e^{\half\alpha_i\cdot(D^T\phi
+D\psi)}\right.\right.\\
\nonumber &&\phantom{2\sum_{i=0}^r\sqrt{m_i}
(\lambda}-\left.\tilde{\cal K}E_{\alpha_i}e^{\half\alpha_i\cdot(D\psi+D^T\phi)}\right)\\
\nonumber &&\phantom{\sum_{i=0}^r\sqrt{m_i}(}-\frac{1}{\lambda}
\left(E_{-\alpha_i}\tilde{\cal K}
e^{\half\alpha_i\cdot D(\phi-\psi)}\right.\\
&&\phantom{2\sum_{i=0}^r\sqrt{m_i}(}-\left.\left.\tilde{\cal K}E_{-\alpha_i}
e^{\half\alpha_i\cdot D^T(\psi -\phi)}\right)\right].
\end{eqnarray}
An alternative possibility might allow \K$(\infty)$ to permute the simple roots,
\begin{equation}\label{}
\nonumber \tilde{\cal K}(\infty)E_{\alpha_i}= E_{\alpha_{\pi(i)}}\tilde{\cal K}(\infty).
\end{equation}
Permutations of this kind are automorphisms of the Ka\v c-Dynkin
diagram for the affine roots. They will not be considered further
in this article.

\p
The next step makes the reasonable assumption
\begin{equation}\label{Kexpansion}
    \tilde{\cal K}=\tilde{\cal K}(\infty)+\frac{k_1}{\lambda}+
    \frac{k_2}{\lambda^2}+\dots \equiv
    1+ \frac{k_1}{\lambda}+\frac{k_2}{\lambda^2}+\dots,
\end{equation}
and examining the terms of order $\lambda^0$ in
\eqref{newKequation} reveals
\begin{equation}\label{}
 \nonumber    {\bf H}\cdot(\nabla_\phi{\cal B}+\nabla_\psi{\cal B})=
 -\sum_{i=0}^r\sqrt{m_i}\, [k_1,E_{\alpha_i}]\, e^{\half\alpha_i\cdot(D^T\phi+D\psi)}.
\end{equation}
Hence,
\begin{equation}\label{kone}
      k_1=\sum_{i=0}^r c_i E_{-\alpha_i},\quad {\cal B}=\sum_{i=0}^rd_i\,
      e^{\half\alpha_i\cdot(D^T\phi+D\psi)} + \tilde{\cal B}(\phi - \psi), \quad c_i=
     \frac{d_i\alpha_i^2}{2\sqrt{m_i}},
\end{equation}
where the defect potential is determined  only partially at this stage, up to a term
which is a function of the defect $\phi -\psi$ (both fields evaluated at $x=0$), and
a set of constants $d_i,\ i=0,\dots, r$.

\p
The order $1/\lambda$ terms in \eqref{newKequation} are more tricky to analyse. They
give
\begin{eqnarray}\label{ktwobitequation}
 \nonumber k_1 {\bf H}\cdot\nabla_\psi{\cal B}+{\bf H}k_1\cdot\nabla_\phi{\cal B}&=&
 -\sum_{i=0}^r\sqrt{m_i}\left(\, [k_2,E_{\alpha_i}]\, e^{\half\alpha_i
 \cdot(D^T\phi+D\psi)}\right.   \\
   && \phantom{\sum_{i=0}^r\sqrt{m_i}}+E_{-\alpha_i}\left.\left[
   e^{\half\alpha_i\cdot D(\phi-\psi)} -e^{\half\alpha_i\cdot D^T(\psi-\phi)}
\right]\right),
\end{eqnarray}
and it is rather clear this splits into two parts, one being an equation for $k_2$,
effectively, and the other involving $\tilde{\cal B}$, the not yet determined part of the
defect potential. The matrix $D$ is not yet determined either, and therefore it is
not yet clear to what extent the exponentials appearing in the second group of
terms in \eqref{ktwobitequation} are independent. However, a sensible first guess for
$\tilde{\cal B}$ is likely to be of the form
\begin{equation}\label{Btildeexpression}
    \tilde{\cal B}=\sum_{i=0}^r\left(p_i \, e^{\half\alpha_i\cdot D(\phi-\psi)}
    +q_i\, e^{\half\alpha_i\cdot D^T(\psi-\phi)}\right),
\end{equation}
on the grounds that $\tilde{\cal B}$ depends only on the defect $\phi-\psi$.
Taking this
to be the case gives
\begin{equation}\label{Btildeequation}
  \sum_{j=0}^r\, \alpha_j\cdot\left( p_j D\alpha_i\,
e^{\half\alpha_j\cdot D(\phi-\psi)} -q_j D^T\alpha_i\,
e^{\half\alpha_j\cdot D^T(\psi-\phi)}\right) =\frac{2\sqrt{m_i}}{c_i}\left(
e^{\half\alpha_i\cdot D(\phi-\psi)} -  e^{\half\alpha_i\cdot D^T(\psi-\phi)}\right).
\end{equation}
For the simplest choice which uses the root data for $a_1^{(1)}$,
$\alpha_1= -\alpha_0=\alpha$ and $D= 1$, the parameters in
\eqref{Btildeexpression} appear in the combinations $p_0+q_1$ and
$p_1+q_0$. To satisfy \eqref{Btildeequation}, $c_0=c_1=c$ and
\begin{equation}\label{aone}
    p_0+q_1=\frac{2\sqrt{m}}{c\alpha^2}=p_1+q_0, \quad m_0=m_1=m.
\end{equation}
Also, from \eqref{kone} it is clear
$d_0=d_1=d=2c\sqrt{m}/\alpha^2$. That is, with the usual
conventions, $p_0+q_1=1/c,\ d_0=d_1=c$. Hence, there is a single
free parameter $c$ and the defect potential reported in
\cite{bczlandau} is recovered. For the case of $a_1^{(1)}$, there
are no further constraints on the parameter $c$ from those
relations arising as coefficients of higher orders in $\lambda$.
However, with other choices, the analysis is less straightforward.

\p First of all, it is clear $D=1$ cannot work since the simple
roots would have to be mutually orthogonal, which is never the
case. An alternative would have to allow cancellation between
different terms on the left hand side of \eqref{Btildeequation}.
If there is a defect and $D+D^T=2$ then cancellation between the
two pieces for a specific $j$-value within the sum cannot happen.
For cancellations to be possible between terms with different
$j$-values it would be necessary to have
\begin{equation}\label{dperm}
    \alpha_jD=-\alpha_{\Pi^{-1}(j)} D^T\equiv -\alpha_j\pi^T D^T,
\end{equation}
where $\Pi$ permutes the simple roots, and $\pi$ achieves the equivalent
via an orthogonal transformation of $\alpha_j$. Since the simple roots
are a linearly independent set, eq\eqref{dperm} requires
\begin{equation}\label{Dtranspose}
    D=-\pi^TD^T=2-D^T
\end{equation}
Hence,
\begin{equation}\label{}
    D^TD=DD^T, \qquad D=2(1-\pi)^{-1},\quad D^T=2(1-\pi^T)^{-1},
\end{equation}
provided $1-\pi$ and $1-\pi^T$ have inverses. These inverses will
fail to exist if $\pi$ leaves a real linear combination of the
simple roots invariant (in other words, $\pi$ has an eigenvector
with eigenvalue $1$). Noting that the two sets of terms in the
putative expression for $\tilde{\cal B}$ \eqref{Btildeexpression}
are the same but ordered differently, it is clear that half the
coefficients are redundant (as was found in the $a_1^{(1)}$ case
described above). Thus a more refined expression is
\begin{equation}\label{newBtildeexpression}
    \tilde{\cal B}=\sum_{i=0}^r s_i \, e^{\half\alpha_i\cdot
    D(\phi-\psi)},
\end{equation}
and \eqref{Btildeequation} will\ be replaced by
\begin{equation}\label{newBtildeequation}
\sum_{j=0}^r\,s_j \, \alpha_j\cdot D\alpha_i\,
e^{\half\alpha_j\cdot D(\phi-\psi)} =\frac{2\sqrt{m_i}}{c_i}\left(
e^{\half\alpha_i\cdot D(\phi-\psi)} -  e^{\half\alpha_{\Pi(i)}\cdot D(\phi-\psi)}\right).
\end{equation}
Clearly, most of the inner products on the left hand side of
\eqref{newBtildeequation} must be zero. An economical set of inner
products which (if possible) will allow \eqref{newBtildeequation}
to be solved are:
\begin{equation}\label{Dcases}
\alpha_j\cdot D\alpha_i=\left\{%
\begin{array}{ll}
    \phantom{-}\alpha_i^2 & \hbox{$j=i$;} \\
    -\alpha_{\Pi(i)}^2 & \hbox{$j=\Pi(i)$;} \\
    \phantom{\alpha_i^2}0 & \hbox{otherwise.} \\
\end{array}%
\right.
\end{equation}
The latter  are achievable provided it is possible to write
\begin{equation}\label{Dalpha}
    D\alpha_i=2\left(\lambda_i-\lambda_{\Pi(i)}\right),
\end{equation}
where the vectors $\lambda_i,\ i=1,\dots ,r$ are the fundamental
weights satisfying
\begin{equation}\label{weights}
    \alpha_i\cdot\lambda_j=\frac{\alpha_i^2}{2}\delta_{ij},
    \quad i,j=1,\dots,r,
\end{equation}
and $\lambda_0\equiv 0$. Note, the necessity to include the term
labelled $0$, creates certain problems. Without it, the Toda theory
would be conformal and eqs\eqref{Dalpha} and \eqref{weights} would
suffice. It is worth noting, since the permutation $\Pi$ preserves
the affine root diagram, that the coefficients $n_i$ in the expression
for $\alpha_0$ in terms of the simple roots satisfy
$$ n_i=n_{\Pi(i)}.$$
Using this fact, it is easy to check that the suggested relation
\eqref{Dalpha} is at the very least consistent since
$$
\sum_{i=0}^r n_i D\alpha_i = 2\sum_{i=0}^r n_i
\left(\lambda_i - \lambda_{\Pi(i)}\right)=2\sum_{i=0}^r n_i
    \lambda_i-2\sum_{i=0}^r n_{\Pi(i)}
    \lambda_{\Pi(i)}\equiv 0.
    $$

\p
As an example, take the case of $a_r^{(1)}$,
whose roots are all of equal length, conventionally chosen to be $\alpha_i^2=2$. Let
\begin{equation}
\nn \alpha_{\Pi(i)}=\alpha_{i+1},\ i=0,\dots, r-1,\quad \alpha_{\Pi(r)}=\alpha_0,
\end{equation}
in which case $\Pi$ is the generator of the elementary cyclic
permutation symmetry of the $a_r^{(1)}$ Ka\v c-Dynkin diagram.
Then,
\begin{equation}\label{Dalphaan}
    D\alpha_i=2\left(\lambda_{i} - \lambda_{i+1}\right), \quad D\alpha_r=2\lambda_r,
    \quad D\alpha_0=-2\lambda_1.
\end{equation}Indeed, for this particular symmetry, $D$ may be written explicitly
\begin{equation}\label{D}
    D=\sum_{i=1}^r\frac{4}{\alpha_i^2}\left( \lambda_i-\lambda_{i+1}\right)\lambda_i^T
    \equiv 2\sum_{i=1}^r \left( \lambda_i-\lambda_{i+1}\right)\lambda_i^T.
\end{equation}
This is consistent with \eqref{matrixrelations} since a direct computation indicates
\begin{equation}\label{}
\nonumber  \left(D+D^T\right)\alpha_k=2\left(2\lambda_k-\lambda_{k+1}-
\lambda_{k-1}\right)\equiv 2\alpha_k.
\end{equation}
 Since the simple roots are linearly
independent, the required property $D+D^T=2$ follows.  Another way
to see this is to note simply
\begin{equation}
\left(D+D^T\right)=2\sum_{i=1}^r\left(2\lambda_i\lambda_i^T-\lambda_{i+1}\lambda_i^T-
\lambda_{i}\lambda_{i+1}^T\right)=2\sum_{i,j=1}^r\lambda_iC_{ij}\lambda_j^T=2,
\end{equation}
where $C_{ij}$\ is the $a_r$ Cartan matrix. In passing, it is also noteworthy
that the antisymmetric matrix $1-D$ is given by the expression
\begin{equation}\label{oneminusD}
1-D=\sum_{i=1}^r\left(\lambda_{i+1}\lambda_i^T-\lambda_i\lambda_{i+1}^T\right),
\end{equation}
which will be required later.

\p
 It is  straightforward to satisfy
\eqref{newBtildeequation} by taking
\begin{equation}\label{ancoeffs}
    c_i=c_{\Pi(i)}=c_{i+1}\equiv c, \quad s_i=s=\frac{2\sqrt{m_i}}{c_i\,\alpha_i^2}\equiv
    \frac{1}{c},
\end{equation}
leaving a single free parameter $c$. In this case, the roots form a single
orbit under $\Pi$, but that may not always be so. (At this stage, it appears free
parameters might be  associated with orbits under $\Pi$.) Using \eqref{ancoeffs}
and \eqref{kone}, the $a_r^{(1)}$ defect potential associated with the
cyclic permutation $\Pi$ is
\begin{equation}\label{anB}
    {\cal B}=\sum_{i=0}^r\left(\, d\, e^{\half\alpha_i\cdot(D^T\phi+D\psi)}+
    \frac{1}{d}\, e^{\half\alpha_i \cdot D(\phi -\psi)}\right).
\end{equation}

\p
The story is not quite finished since there remain other pieces to be checked. The
remaining terms in \eqref{ktwoequation}
 at order $1/\lambda$ need to be examined to determine $k_2$. Since the exponential
 factors are expected generally to be different the terms that need to match are
 \begin{eqnarray}\label{ktwoequation}
   \nonumber -2\sqrt{m_i}\, [k_2,E_{\alpha_i} ]
   &=& d_i\left(\alpha_i
   \cdot Dk_1{\bf H}+\alpha_i\cdot D^T{\bf H}k_1\right) \\
 \nonumber   &=&  d_i\left(\alpha_i\cdot(D+D^T){\bf H}k_1 +\alpha_i\cdot
   D\,[k_1,{\bf H}]\right) \\
 \nonumber   &=&  d_i\left(2\alpha_i\cdot{\bf H}k_1 +\alpha_i\cdot D
    \sum_{j=0}^r c_j\alpha_j E_{-\alpha_j}\right)\\
 \ &=& d_i\left(2\alpha_i\cdot{\bf H}k_1 +c_i\alpha_i^2
 E_{-\alpha_i}-c_{\Pi^{-1}(i)}\alpha_{\Pi^{-1}(i)}^2E_{-\alpha_{\Pi^{-1}(i)}}\right),
 \end{eqnarray}
where the expression \eqref{kone} for $k_1$ has been used along with the facts
\eqref{Dcases}. For the choice of roots corresponding to $a_r^{(1)}$ with
the permutation and the conventions used above, eq\eqref{ktwoequation}
simplifies to
\begin{equation}\label{arktwoequation}
-[k_2,E_{\alpha_i} ]=d^2\left(\alpha_i\cdot{\bf H}\sum_{j=0}^r E_{-\alpha_{j}} +
 E_{-\alpha_i}-E_{-\alpha_{i-1}}\right).
\end{equation}
The natural algebra grading dictates that $k_2$ has the form
\begin{equation}\label{}
    k_2=\sum_{k,l=0}^r c_{kl} E_{-\alpha_k}E_{-\alpha_l},
\end{equation}
and therefore
\begin{equation}\label{arktwoexpression}
-[k_2,E_{\alpha_i}]=\alpha_i\cdot{\bf H}\sum_{k=0}^r(c_{ki}+c_{ik})E_{-\alpha_k}
+\sum_{k=0}^r c_{ki}\, (\alpha_i\cdot\alpha_k)\, E_{-\alpha_k}.
\end{equation}
Hence, comparing with \eqref{arktwoequation} yields
\begin{equation}\label{ktwocomponents}
c_{ki}+c_{ik}=d^2,\quad c_{ii}=\frac{d^2}{2},\quad c_{i-1\, i}=d^2,\quad
c_{i+1\, i}=0,\quad i,k=0,
\dots, r
\end{equation}
which allows for some considerable freedom in $k_2$.

\p
On the other hand, $k_2\equiv 0$
when evaluated in an $r+1$ dimensional representation for which
\begin{equation}\label{}
    (E_{-\alpha_i})_{ab}=\delta_{ai}\delta_{b\, i-1}, \quad a,b =1,\dots, r+1.
\end{equation}
Indeed, in that particular representation it can be checked directly that
a complete expression for $\tilde{\cal K}$ is given by
\begin{equation}\label{rplusoneK}
\tilde{\cal K}=1+\frac{d}{\lambda}\sum_{i=0}^r\, E_{-\alpha_i}.
\end{equation}

\p
In general, at the next order, $1/\lambda^2$, there may be further constraints
on $k_2$. To examine these
it is necessary to analyse the following
\begin{eqnarray}\label{oneoverlsquared}
  \nn  k_2 {\bf H}\cdot\nabla_\psi{\cal B}+{\bf H}k_2\cdot\nabla_\phi{\cal B}&=&
 -\sum_{i=0}^r\sqrt{m_i}\left(\, [k_3,E_{\alpha_i}]\, e^{\half\alpha_i
 \cdot(D^T\phi+D\psi)}\right.   \\
   && \phantom{\sum_{i=0}^r\sqrt{m_i}}+\left.\left(E_{-\alpha_i}k_1
   -k_1 E_{-\alpha_{\Pi^{-1}(i)}}\right)e^{\half\alpha_i\cdot D(\phi-\psi)} \right),
\end{eqnarray}
where, in the last term on the right hand side, the relation \eqref{dperm}
has been used together with a reordering of the sum. Apart from an equation for
$k_3$ there is a further equation involving $k_2$ by itself, which might, in
principle, require further restrictions. Thus, inserting the expression for
${\cal B}$,
\begin{equation}\label{}
    \nn \frac{1}{2d}[k_2,{\bf H}]\cdot D^T\alpha_i=\sqrt{m_i}\left(
    E_{-\alpha_i} k_1 - k_1 E_{-\alpha_{\Pi^{-1}(i)}}\right),
\end{equation}
which can be shown to be identically satisfied using the already obtained expressions
for $k_1$ and $k_2$, and using information concerning the Lie algebra of $A_r$.
Specifically, it is important to note that the commutator $[E_{-\alpha_k},E_{-\alpha_l}]$
vanishes except when $k=l\pm 1$. So, perhaps surprisingly, there is actually no additional
constraint on $k_2$.

\p
On the other hand, the equation for $k_3$ leads to
\begin{equation}\label{kthree}
 \nn   k_3=\sum_{k,l,m=0}^r c_{klm}E_{-\alpha_k}E_{-\alpha_l}E_{-\alpha_m},
\end{equation}
which is of the expected form with respect to the grading alluded to above,
with a set
of relations among the coefficients. These are quite complicated and summarised below
\begin{eqnarray}\label{kthreecomponents}
\nn & c_{ijk}+c_{ikj}+c_{jik}+c_{jki}+c_{kij}+c_{kji}=d^3,  &\quad i,j,k=0,\dots, r;\\
\nn & c_{ijk}+c_{jik}+c_{jki}=dc_{jk},&\quad i,j=0,\dots,r,\ k=j\pm 1; \\
\nn &\\
\nn & c_{k\, i-1\, i} + c_{i-1\, k\, i}+c_{i-1\, i\, k}=d^3,&\quad k\ne i+1,i,i-1,i-2;\\
\nn &c_{i+1\, k\, i}+c_{i+1\, i\, k}+c_{k\, i+1\, i}=0,&\quad k\ne i+2,i+1,i,i-1;
\end{eqnarray}
\begin{eqnarray}
\nn & c_{i-1\, i\, i-1}+2c_{i-1\, i-1\, i}=d^3, \ \ c_{i+1\, i\, i+1}+
2c_{i+1\, i+1\, i} =0; & \\
\nn & c_{i\, i-1\, i-1}-c_{i-1\, i-1\, i}=-\frac{d^3}{2}=c_{i+1\, i+1\, i}-
c_{i\, i+1\, i+1},&\quad i=0,\dots, r;\\
\nn &\\
\nn &  c_{i-2\, i-1\, i}=d^3=2c_{i-2\, i-1\, i}+c_{i-2\, i\, i-1};&\\
\nn & 2c_{i+2\,i-1\,i}+2c_{i-2\, i+1\, i}+c_{i+1\, i\, i-1}+c_{i-1\, i\, i+1}=d^3;&\\
 & c_{i+2\, i+1\, i}=0=c_{i-1\, i-2\, i}+c_{i-1\, i\, i-2},&\quad i=0,\dots, r.
\end{eqnarray}
These relations refer to the general case $a_r$. However, for the special case of
$a_2$, because of the identification of the labelling modulo $r+1$, some of the relations
are modified and do not fit the general type. Also, these relations do not specify
the coefficients of $k_3$ uniquely; there is further freedom.

\p
We have no reason to suppose there is any problem, at least in principle, in
continuing this iterative process to determine $\tilde{\cal K}$, but it does not
lead easily to a closed form, and such an expression has not been
found by alternative means. It remains an open question.

\section{Momentum revisited}
\label{s:momentum}

In section [\ref{s:fsf}] it was pointed out that a defect can
allow momentum to be conserved provided the momentum density is
suitable modified. There, the example used was a pair of free
scalar fields with a single defect at $x=0$. In this section the
possibility of defining a conserved momentum within the more
general situation will be explored. The starting point is
eq\eqref{Pdot} in which each scalar field is replaced by a
multi-component field and the products interpreted accordingly.
Thus, using the defect conditions \eqref{affineconditions}
\begin{eqnarray}\label{Pmess}
\nn  \frac{dP}{dt}&=&\half\left[\left(\partial_{t}\phi\right)^{2}+
\left(E\partial_{t}\phi+D\partial_{t}
\psi-\nabla_{\phi}{\cal B}\right)^{2}-2V(\phi)\right] \\
&&-\half\left[\left(\partial_{t
}\psi\right)^{2}+\left(D^{T}\partial_{t}\phi-F\partial_{t}
\psi+\nabla_{\psi}{\cal B}\right)^{2}-2W(\psi)\right],
\end{eqnarray}
where all fields on the right hand side are evaluated at $x=0$. In order that
\eqref{Pmess} be the total time derivative of a functional of the fields $\psi$
and $\phi$ evaluated at $x=0$, the following relations must hold
\begin{eqnarray}\label{Pconditions}
 \nn  \partial_t{\phi}\cdot (1+ E^T E- D D^{T})\partial_t{\phi}&=&0 \\
\nn  \partial_t{\psi}\cdot (1-D^TD -F^T F)\partial_t{\psi}&=&0 \\
\nn  \partial_t{\phi}\cdot ( E^TD+ D F)\partial_t{\psi} &=&0\\
\half(\nabla _{\phi}{\cal
B})^{2} -\half(\nabla _{\psi}{\cal
B})^{2}&=&V(\phi)-W(\psi).
\end{eqnarray}
The first three of the required relations \eqref{Pconditions} follow
directly from the already derived relationships between $E,F$ and $D$,
\eqref{matrixrelations}; the fourth requires a more careful examination
but also follows from properties of the matrix $D$.

\p
Assembling the pieces of the defect potential contained in eqs(\ref{kone},
\ref{newBtildeexpression}),
\begin{equation}\label{B}
    {\cal B}=\sum_{i=0}^{r}\left[d_{i}e^{\frac{1}{2}\alpha_{i}\cdot(D^{T}\phi+D\psi)}
+s_{i}e^{\frac{1}{2}\alpha_{i}\cdot D(\phi-\psi)}\right]\equiv\sum_{i=0}^{r}
\left(f_{i}+g_{i}\right),
\end{equation}
and evaluating the left hand side of the fourth of eqs\eqref{Pconditions}
gives
\begin{eqnarray}\label{fourthcheck}
  \half(\nabla _{\phi}{\cal
B})^{2} -\half(\nabla _{\psi}{\cal
B})^{2} &=& \half\sum_{i,j=0}^r d_is_j\, \alpha_i\cdot D\alpha_j\, e^{\half \alpha_i\cdot
D(\phi-\psi)+\half\alpha_j\cdot(D^T\phi + D\psi)}.
\end{eqnarray}
Using the basic properties of $D$ expressed in eqs(\ref{Dcases}, \ref{dperm}), the
right hand side of \eqref{fourthcheck} may be rewritten
\begin{eqnarray}\label{fourthcheckagain}
 \nn &&\half\sum_{j=0}^r s_jd_j\alpha_j^2\, e^{\half\alpha_j\cdot(D+D^T)\phi}
  -\half\sum_{j=0}^r s_{\Pi(j)}d_j\alpha_{\Pi(j)}^2\, e^{\half\alpha_{\Pi(j)}\cdot D(\phi
   -\psi)+\half\alpha_j\cdot(D^T\phi+D\psi)}\\
 \nn & &\ \ \ \ \ \ \ \ \ =\half\sum_{j=0}^r s_jd_j\alpha_j^2\, e^{\alpha_j\cdot\phi} -
  \half\sum_{j=0}^r s_{\Pi(j)}d_j\alpha_{\Pi(j)}^2\, e^{\alpha_j\cdot\psi} =
V(\phi)-V(\psi)
\end{eqnarray}
where $V$ is the affine Toda potential, provided, for each $j=0,\dots, r$,
\begin{equation}\label{}
 \nn   \half s_jd_j\alpha_j^2=n_j= \half s_{\Pi(j)}d_j\alpha_{\Pi(j)}^2.
\end{equation}
Since $\Pi$ is a symmetry of the extended root diagram, $\alpha_j^2=\alpha_{\Pi(j)}^2$,
and $s_j=s_{\Pi(j)}=n_j/d_j.$

\p
Finally, similar manipulations lead to the conclusion
\begin{equation}\label{}
    \frac{dP}{dt}=\frac{dU}{dt},\quad U=-\sum_{i=0}^r d_ie^{\half\alpha_i\cdot(D^T\phi+D\psi)}
    +\left.\sum_{i=0}^r s_i e^{\half\alpha_i\cdot D(\phi-\psi)}\right|_{x=0}.
\end{equation}
Hence, the generalised conserved momentum may be written
\begin{equation}\label{newmomentum}
    {\cal P}=P-U=P+\sum_{i=0}^r(f_i-g_i).
\end{equation}
This is to be compared with the standard expression for the conserved energy
\begin{equation}\label{energy}
    {\cal E}=E+{\cal B}=E+\sum_{i=0}^r(f_i+g_i),
\end{equation}
and it is worth noting that the `light-cone' versions of these\begin{equation}\label{eandp}
    {\cal P}_\pm =\half ({\cal E}\pm {\cal P})
\end{equation}involve only the $f_i$ or $g_i$, respectively, meaning (because of the
relationship between $d_i$ and $s_i$) that a Lorentz
transformation of the bulk quantities can be compensated by a
change of scale in $d_i$, just as was found in the linear case
\eqref{EplusorminusP}. If there are several defects located at
different places the total energy and the total modified momentum
will be conserved with contributions arising at each defect.

\section{Higher spin conserved charges}

Given the existence of a Lax pair incorporating the defect it is
certainly expected that there will be an infinite number of
conserved charges whose generating function will be of the form
\cite{Bow95},
\begin{equation}\label{chargegenerator}
  {\cal Q}(\lambda)  ={\rm tr}\left(\exp{\left[\int_{-\infty}^a dx\,
  \widehat{a}^{\, <}_x(\lambda)\right]}\
    {\cal K(\lambda)}
   \ \exp{\left[\int_b^\infty dx\, \widehat{a}^{\, >}_x(\lambda)\right]}\right).
\end{equation}
The quantity ${\cal Q}(\lambda)$ is time-independent because of
the zero curvature condition satisfied by the two gauge
fields $\widehat{a}^{\, <}_x$ and $\widehat{a}^{\, >}_x$, and because of the
condition on ${\cal K}$ expressed by
\eqref{overlapgauge}. Expanding ${\cal Q}(\lambda)$ as a Laurent
series in $\lambda$ reveals the conserved quantities as the
coefficients of the various powers of $\lambda$ that occur in the
expansion. On the other hand, it has not yet proved possible to
calculate the conserved charges from the Lax pair despite it being
feasible in principle. In fact, there is a subtlety since not all
charges that are conserved in the bulk survive in the
presence of a defect - indeed that might be expected given the experience gained
by dealing with boundaries. There, typically all `momentum-like' charges are
lost although `energy-like' charges are preserved, albeit   modified. In the
presence of a defect it was demonstrated in the previous section
that momentum and energy themselves
(spin $s=\pm 1$) are both preserved, at least after suitable adjustments have been made
to include a contribution to the momentum from the defect.
A glance at the expression for \K\ provided by \eqref{rplusoneK}
suggests there is a lack of balance between $\lambda$ and $1/\lambda$ (or equivalently
$d$ and $1/d$) which might indicate that charges of positive spins
should behave generally quite differently to those of negative spin.
To illustrate this, it is enough to examine the charges which would have spin $s=\pm 2$
in the bulk.

\p
A general ansatz for a spin $s=\pm 3$ bulk density $T_{\pm 3}$ (using light-cone coordinates
$x^{\pm}=(t\pm x)/\sqrt{ 2}$) reads (see \cite{Bow95})
\begin{equation}\label{Tthree}
T_{\pm 3} = \frac{1}{ 3} A_{abc} \partial_\pm \phi_a \partial_\pm \phi_b \partial_\pm \phi_c +
B_{ab} \partial_\pm^2 \phi_a \partial_\pm \phi_b,
\end{equation}
where the coefficients $A_{abc}$ are completely symmetric and the coefficients
$B_{ab}$ are antisymmetric. An explicit calculation reveals that  the continuity
equation
\begin{equation}
\nn \partial_\mp T_{\pm 3}=\partial_\pm \Theta_{\pm 1}
\end{equation}
is satisfied for the choice
\begin{equation}\label{Theta}
{\Theta_{\pm 1}=-{1\over 2} B_{ab}\partial_\pm\phi_aV_b,}
\end{equation}
provided
\begin{equation}\label{ABequation}
A_{abc}V_a+B_{ab}V_{ac}+B_{ac}V_{ab}=0,
\end{equation}
where
\begin{equation}
\nn V_b=\frac{\partial V}{\partial\phi_b},\qquad V_{bc}=
\frac{\partial^2 V}{\partial\phi_b\partial\phi_c}.
\end{equation}
As explained in \cite{Bow95} the latter set of equations determine $A_{abc}$ and
$B_{ab}$ up to an overall multiplicative constant.
It is often convenient to write \eqref{ABequation} in a field independent way,
in which case for each simple root $\alpha_i$ (or $\alpha_0$)
\begin{equation}\label{ABrootequation}
    A_{abc}(\alpha_i)_c+B_{cb}(\alpha_i)_c(\alpha_i)_a + B_{ca}(\alpha_i)_c(\alpha_i)_b=0.
\end{equation}
Then, on careful inspection \eqref{ABrootequation}
requires $B_{ab}$ to be a constant multiple of the right hand side of \eqref{oneminusD}
 and, therefore, without losing generality,
it is convenient to write
\begin{equation}\label{BDequation}
B=1-D.
\end{equation}

\p
Light-cone coordinates are not convenient for models with a defect but a
suitable rearrangement of \eqref{Tthree} is
\begin{equation}\label{newTthree}
    \partial_t(T_{\pm 3} - \Theta_{\pm 1})=\pm \partial_x (T_{\pm 3} + \Theta_{\pm 1}),
\end{equation}
leading to candidate bulk contributions to spin 2 conserved charges of the form
\begin{equation}\label{}
\nn    Q_{\pm 2}=\int_{-\infty}^0\, dx \, (T^\phi_{\pm 3} - \Theta^\phi_{\pm 1}) +
    \int_{0}^\infty\, dx\, (T^\psi_{\pm 3} - \Theta^\psi_{\pm 1}).
\end{equation}
Then, the time derivatives of the candidate charges will be given by terms evaluated at
the position of the defect
\begin{equation}\label{Qtwodot}
    \frac{dQ_{\pm 2}}{dt}=\pm \left[T^\phi_{\pm 3} + \Theta^\phi_{\pm 1}\right]_0
    \mp\left[T^\psi_{\pm 3} + \Theta^\psi_{\pm 1}\right]_0.
\end{equation}
Provided the right hand side of \eqref{Qtwodot} may be written as the time derivative
of a functional, ${\cal B}_{\pm 2}$, depending upon the two fields $\phi$ and $\psi$
evaluated at $x=0$, the quantities
$$\widehat Q_{\pm 2}=Q_{\pm 2} -{\cal B}_{\pm 2}(\phi(0,t),\psi(0,t))$$
will be conserved.

\p
To facilitate implementing the defect conditions in \eqref{Qtwodot} it is useful to
reorganise \eqref{affineconditions} to read as follows
\begin{eqnarray}\label{newaffineconditions}
\nn  \sqrt{2}\partial_-\phi &=& D\partial_t(\phi-\psi) +\nabla_\phi {\cal B}, \\
 \sqrt{2}\partial_-\psi &=&  D^T\partial_t(\psi-\phi) -\nabla_\psi {\cal B},\\
\label{newaffineconditionsplus}\nn \sqrt{2} \partial_+\phi &=& D^T\partial_t
\phi+D\partial_t\psi -\nabla_\phi {\cal B}, \\
 \sqrt{2} \partial_+\psi &=& D\partial_t\psi+D^T\partial_t\phi +\nabla_\psi {\cal B}.
\end{eqnarray}
Notice that the second pair also implies
\begin{equation}
\nn \sqrt{2}\partial_+(\phi-\psi)=-(\nabla_\phi{\cal B}+\nabla_\psi{\cal B})
\end{equation}
independently of the matrix $D$.

\p
The two different groupings are equivalent but convenient since the boundary conditions
will be used to replace light-cone derivatives evaluated at the defect. It is quite
complicated to implement these conditions in \eqref{Qtwodot}. However, before
examining the full detail it is soon
apparent there is a distinction between the two spins. To see this, examine the
highest order terms in the time derivatives of the fields appearing on the right hand
side of \eqref{Qtwodot} after the defect conditions have been implemented. For example,
in the case of spin -2 a typical one of these up to a numerical factor is
\begin{equation}\label{typicalmess}
\partial_t\phi_p\partial_t\phi_q\partial_t\phi_r A_{abc}(D_{ap}D_{bq}D_{cr}+
D_{pa}D_{qb}D_{rc}).
\end{equation}
Clearly, because of \eqref{Dtranspose} the expression \eqref{typicalmess} does not
vanish and it cannot be written
as a total time derivative. Hence the candidate charge $Q_{-2}$ cannot be modified
in any simple way to yield a conserved quantity. On the other hand, this particular type of
term vanishes identically for the spin +2 case. Thus, at best $Q_2$ may have a conserved
modification, and the next part of this section will explore that possibility in detail.
The key lies in the fact that the same combination of derivatives of fields $\phi$ and $\psi$
 appears on the right hand sides of the `plus' parts of both of the boundary conditions
 supplied by \eqref{newaffineconditionsplus}.

\p
To facilitate checking it is useful to make a small change of notation, abbreviating the
combination $D^T\phi +D\psi$ to $R$ and $(\nabla_\phi{\cal B})_a$ to ${\cal B}^\phi_a$,
$(\nabla_\phi V)_a$ to $V^\phi_a$, etc. Then, the terms which need to be examined are the
following
\begin{eqnarray}
\nn  && \frac{1}{6\sqrt{2}}A_{abc}\left[(\partial_t R_a-{\cal B}^\phi_a)(
\partial_t R_b-{\cal B}^\phi_b)(\partial_t R_c-{\cal B}^\phi_c)-
(\partial_t R_a-{\cal B}^\psi_a)(
\partial_t R_b-{\cal B}^\psi_b)(\partial_t R_c-{\cal B}^\psi_c)\right]\\
 \nn  &+& \frac{1}{2\sqrt{2}}B_{ab}\left[(2\partial_t^2 R_a -2\partial_t{\cal B}^\phi_a
   + V^\phi_a)(\partial_t R_b -{\cal B}^\phi_b)-
   (2\partial_t^2 R_a +2\partial_t{\cal B}^\psi_a
   +V^\psi_a)(\partial_t R_b +{\cal B}^\psi_b)\right]\\
   &-& \frac{1}{2\sqrt{2}}B_{ab}\left[(\partial_t R_a-{\cal B}^\phi_a)V_b^\phi-
(\partial_t R_a+{\cal B}^\psi_a)V_b^\psi\right].
\end{eqnarray}
Clearly, as already noted, all the terms with three time derivatives cancel.

\p
The terms with two time derivatives, after making use of the explicit form of the defect
potential and the basic property \eqref{ABrootequation}, are seen to be a total
time derivative
\begin{equation}\label{twoderivative}
    -\frac{1}{\sqrt{2}}\partial_t\, \left[B_{ab} \partial_tR_a({\cal B}^\phi_b+
    {\cal B}^\psi_b)\right ].
\end{equation}
The terms with a single time derivative gather together to
\begin{eqnarray}
 \nn  & & \frac{1}{2\sqrt{2}}\,A_{abc}\partial_tR_a({\cal B}^\phi_b-{\cal B}^\psi_b)
({\cal B}^\phi_c +{\cal B}^\psi_c)
   +\frac{1}{\sqrt{2}}\,B_{ab}(\partial_t{\cal B}^\phi_a{\cal B}^\phi_b-
  \partial_t{\cal B}^\psi_a{\cal B}^\psi_b +(V_a^\phi-V_a^\psi)\partial_tR_b),
\end{eqnarray}
which is also a total derivative
\begin{equation}\label{}
    \partial_t\left(\frac{1}{4\sqrt{2}}\sum_{i,j}\alpha_i\cdot D(1-D)\alpha_j
    \, e^{\frac{1}{2}\alpha_i\cdot D(\phi-\psi)}e^{\frac{1}{2}\alpha_j\cdot R}\right).
\end{equation}
This follows, on using \eqref{ABrootequation}, \eqref{BDequation}
and the basic defining property of $D$, \eqref{dperm}.

\p
The terms without any time derivatives are
\begin{equation}\label{}
    -\frac{1}{6\sqrt{2}}\, A_{abc}({\cal B}^\phi_a{\cal B}^\phi_b{\cal B}^\phi_c -
    {\cal B}^\psi_a{\cal B}^\psi_b{\cal B}^\psi_c )-\frac{1}{\sqrt{2}}\, B_{ab}(
    V_a^\phi{\cal B}^\phi_b +V_a^\psi{\cal B}^\psi_b);
\end{equation}
using \eqref{ABrootequation}, the explicit expression for the defect potential,
and the defining property of $D$ these are shown to exactly cancel. Hence, a
suitably modified spin 2 charge is conserved.

\p
For the example we have been using, the spin +2 charge is preserved, the spin -2 is lost.
On the other hand, a moment's thought reveals that a slightly different example would
interchange the r\^oles of these two charges. Return to the equation defining $\tilde{\cal K}$,
\eqref{Kexpansion}. To solve this it was supposed $\tilde{\cal K}$ had an expansion in
inverse powers of the spectral parameter $\lambda$. If instead $\tilde{\cal K}$ was presumed to
have an expansion in positive powers of $\lambda$, the properties of $D$, the
relationships between $D,\ E$ and $F$ and the
boundary potential would be slightly different but another consistent solution would be
obtained. That is,
\begin{equation}\label{}
    E=F=1+D^T, \quad D+D^T=-2,
\end{equation}
with
\begin{eqnarray}\label{alternativeaffineconditions}
\nn  \sqrt{2}\partial_-\phi &=& -D^T\partial_t\phi - D\partial_t\psi +\nabla_\phi {\cal B}, \\
 \sqrt{2}\partial_-\psi &=& - D^T\partial_t\phi -D \partial_t\psi -\nabla_\psi {\cal B},\\
\label{newaffineconditionsplus}\nn \sqrt{2} \partial_+\phi &=& -
D\partial_t(\phi-\psi) -\nabla_\phi {\cal B}, \\
 \sqrt{2} \partial_+\psi &=& -D^T\partial_t(\psi -\phi) +\nabla_\psi {\cal B},
\end{eqnarray}
and $D$ is replaced by $-D$ in the defect potential. For this
solution, the spin -2 charge is conserved instead. This means that
the presence of two defects of opposing types located at different
positions would entail a complete loss of the spin two charges.
Since even spin charges are able to distinguish between particles
and antiparticles in the $a_r^{(1)}$ affine Toda field theory, the
distinction between particles and antiparticles should presumably
be lost in a typical situation with many defects. A full
discussion of conserved charges would require deducing the form
they take from \eqref{chargegenerator} and that has not yet been
done. It is conceivable the apparently broken charges are repaired
by assembling combinations with several different spins, but that
remains to be seen. Such a possibility is certainly feasible since the
defect explicitly breaks Lorentz invariance. In the presence of a
boundary, something similar needed to be done since the `energy-like'
combinations of positive and negative spins  have to be used to assemble
conserved quantities.

\subsection{Defect conditions from B\"acklund Transformations}

The argument of section 3 aims to constrain the boundary
Lagrangian by imposing the condition that the resulting boundary
conditions lead to the existence of a gauge transformation between
the two zero-curvature Lax pairs on each side of the defect. In
this case, one is guaranteed the classical integrability of the
model. In the case of sine-Gordon theory, the resulting boundary
conditions are exactly the same as the auto-B\"acklund
transformation for sine-Gordon theory. In some ways this is a
remarkable result, since the B\"acklund transformations are
equations which are true for all $x$,  the
corresponding boundary conditions could be imposed at any value of
$x$ without
changing the value of the fields far away from the boundary.
Indeed the form of the boundary potential determines the boundary
conditions, which assumed true for all $x$ can be
cross-differentiated to determine the bulk equations of motion.

\p
The purpose of this section is to work from the opposite point of view, and
ask when boundary conditions of the form \eqref{affineconditions}, which arise from a
boundary Lagrangian coupling two sets of $n$ scalar fields $\psi,$\ $\phi$
can be interpreted as a B\"acklund transformation leading to Lorentz
invariant second order equations for the fields in the bulk. Under
some fairly mild assumptions the solutions for $D,$\ $E$
and $F$ found in section 3 will be recovered.

\p
For these purposes it will be convenient to think of $\phi$ and $\psi$ as
components of a $2n$ dimensional vector $\Psi=(\phi,\psi)$. In this
notation the boundary condition can be written
\begin{equation}
\partial_{x}\Psi=M\partial_{t}\Psi+\left(\begin{array}{c}
-{\cal B}_{\phi}\\
{\cal B}_{\psi}\end{array}\right),\end{equation}
where \begin{equation}
M=\left(\begin{array}{cc}
E\phantom{^T} & \phantom{-}D\\
D^{T} & -F\end{array}\right).\end{equation}
 Differentiating with respect to $x$ and $t$ and removing the cross-derivatives
reveals\begin{equation}
\partial_{x}^{2}\Psi=M^{2}\partial_{t}^{2}\Psi+\left( MN+NM\right)
\partial_{t}\Psi+N\left(\begin{array}{c} -{\cal B}_{\phi}\\
{\cal B}_{\psi}\end{array}\right),\end{equation}
where \begin{equation}
N=\left(\begin{array}{cc}
\phantom{}-{\cal B}_{\phi\phi} & -{\cal B}_{\phi\psi}\\
\phantom{-}{\cal B}_{\psi\phi} & \phantom{-}{\cal B}_{\psi\psi}
\end{array}\right).\end{equation}
Demanding  the resulting equations for $\phi$  and $\psi$ be decoupled,
and Lorentz-invariant, leads immediately to
\begin{eqnarray}
\label{Msquare} M^{2} & = & 1\\
\label{MN} MN+NM & = & 0\\
\label{BVW}{\cal B}_{\phi}^{2}-{\cal B}_{\psi}^{2} & = & 2(V(\phi)-W(\psi)),\end{eqnarray}
where $V$ and $W$ should be the bulk potentials. The next step is to solve
these three equations in turn. In components, \eqref{Msquare} is equivalent to
\begin{equation}\label{EDF}
E^{2}+DD^{T}=1,\quad F^{2}+D^{T}D=1,\quad ED-DF=0,\quad D^TE-FD^T=0.\end{equation}
For a real boundary potential, $E$ and $F$ may be taken to be real
antisymmetric matrices. So, the first of these equations may be rewritten
 \begin{equation}
DD^{T}=1-E^{2}=(1-E)(1-E)^{T},\end{equation}
while the last of equations \eqref{EDF} are the same.
But, if $E$ is real and antisymmetric, it has purely imaginary eigenvalues
and hence $(1-E)$ is invertible. Then \begin{equation}
(1-E)^{-1}D((1-E)^{-1}D)^{T}=1,\end{equation}
indicating that $D=(1-E)R$ where $R\in O(n)$. By making an
orthogonal transformation on, say, the basis of $\psi$, $R$ may
be replaced by the unit matrix, and
the relationship $D=1-E$ is recovered. It follows $D$ is also
invertible and from this it is easy to prove $F=E=1-D$.

\p
Eq\eqref{MN} is a second order
linear differential equation for ${\cal B}$, which simplifies
provided a different basis corresponding to linear combinations of $\phi$ and
$\psi$ is used.  Choosing variables
\begin{equation}
\begin{array}{ccc}
\mu & = & D^T\phi+D\psi\\
\nu & = & -D(\phi-\psi),\end{array}\end{equation}
and making the change to this basis, noting
\begin{equation}
\begin{pmatrix}
\partial_\phi\\
\partial_\psi\\
\end{pmatrix}=
\begin{pmatrix}
D\phantom{^T}&-D^T\\
D^T&\phantom{-}D^T\\
\end{pmatrix}
\begin{pmatrix}
\partial_\mu\\
\partial_\nu\\
\end{pmatrix},
\end{equation}
 the condition \eqref{MN} becomes
\begin{equation}
2\left(\begin{array}{cc}
-D\phantom{^T} & -D^T\\
-D^{T} & \phantom{-}D^{T}\end{array}\right)\left(\begin{array}{cc}
0 & {\cal B}_{\mu\nu}\\
{\cal B}_{\nu\mu} & 0\end{array}\right)\left(\begin{array}{cc}
\phantom{-}D^T & D\\
-D\phantom{^T} & D\end{array}\right)=0,\end{equation}
which is solved if ${\cal B}_{\mu\nu}=0$, with solution
${\cal B}={\cal B}_{1}(\mu)$+${\cal B}_{2}(\nu)$.

\p
Finally, \eqref{BVW} is also relatively neat in terms of the
variables $\mu$ and $\nu$ and can be written\begin{equation}
{\cal B}_{\phi}^{2}-{\cal B}_{\psi}^{2}=-4\,D^T_{\alpha\beta}\,
\frac{\partial {\cal B}_{1}}{\partial\mu_{\alpha}}\frac{\partial
{\cal B}_{2}}{\partial\nu_{\beta}}=
2(V(\phi)-W(\psi)).\end{equation}
Bearing in mind the expected forms of $V$ and $W$, it is convenient to take,
\begin{equation}
{\cal B}_{1}=\sum_{i}c_{i}e^{v^{(i)}\cdot\mu},\quad{\cal B}_{2}=
\sum_{j}d_{j}e^{w^{(j)}\cdot\nu}.\end{equation}
 Since $\phi=\frac{1}{2}(\mu-\nu),$   a non-zero
$V(\phi)$ requires $v^{(i)}=-w^{(j)}$ for some $i,j$. By a suitable choice of
ordering, it is convenient to arrange that $v^{(i)}=-w^{(i)}.$ Now,
if  the vector space spanned by $\phi$ is denoted by $S_{\phi}$, and
the vector space spanned by $\psi$ is denoted by $S_{\psi}$, then
\eqref{BVW} implies that for each $i$ and $j$ one of the following
three possibilities must hold
\begin{equation}\label{vD}
\begin{array}{rcl}
 v^{(i)}\cdot\mu-v^{(j)}\cdot\nu & \in & S_{\phi}\\
 v^{(i)}\cdot\mu-v^{(j)}\cdot\nu & \in & S_{\psi}\\
D^T_{\alpha\beta}v_{\alpha}^{(i)}v_{\beta}^{(j)} & = & 0.\end{array}\end{equation}
 The first condition holds if $i=j$, and indeed it can be shown that
it holds only in this case. To see this, construct a contradiction. Suppose
$v^{(i)}\cdot\mu-v^{(j)}\cdot\nu\in S_{\phi},$ then
\begin{equation}
\nn v^{(i)}\cdot\mu-v^{(j)}\cdot\nu-(v^{(j)}\cdot\mu-v^{(j)}\cdot\nu)=
(v^{(i)}-v^{(j)})\cdot\mu\in S_{\phi}.\end{equation}
But, by the definition of $\mu$,   \begin{equation}
\nn (v^{(i)}-v^{(j)})\cdot (\mu- D^T\phi)=(v^{(i)}-v^{(j)})\cdot D\psi.\end{equation}
The left-hand side of this equation is in $S_{\phi}$ but the right
hand side cannot be in $S_{\phi}$, and indeed cannot vanish since $i\neq j$
and $D$ is invertible.

\p
By similar considerations, for fixed $i$ the second of the conditions \eqref{vD} can
only be satisfied for at most one $j$ which will not be equal to
$i$. Again, choosing labels appropriately,  this
condition is taken to hold for $j=i+1.$ If it is also supposed  this process
terminates and that for some $N$, $v^{(N+1)}=v^{(1)},$ then
\begin{equation}\label{vmn}
\sum_{i=1}^{N}\left(v^{(i)}\cdot\mu-v^{(i)}\cdot\nu\right)=\sum_{i=1}^{N}
\left(v^{(i)}
\cdot\mu-v^{(i+1)}\cdot\nu\right).\end{equation}
The left-hand side of \eqref{vmn} is clearly in $S_{\phi}$ whilst the right is
in $S_{\psi}$ and thus both sides must vanish. This in turn implies
that \begin{equation}
\sum_{i=1}^{N}v^{(i)}=0.\end{equation}

\p
Finally let us turn to the third of conditions \eqref{vD}. This
condition should hold whenever the first two do not, that is, if $i\neq j$
and $i+1\neq j.$ If $|i-j|>1$ then
\begin{equation}
\begin{array}{rcl}
D^T_{\alpha\beta}v_{\alpha}^{(i)}v_{\beta}^{(j)} & = & 0\\
D^T_{\alpha\beta}v_{\alpha}^{(j)}v_{\beta}^{(i)} & = &
D_{\alpha\beta}v_{\alpha}^{(i)}v_{\beta}^{(j)}=0,\end{array}\end{equation}
 implying that $v^{(i)}\cdot v^{(j)}=0$ and
$E_{\alpha\beta}v_{\alpha}^{(i)}v_{\beta}^{(j)}=0$. Next, consider the remaining possibility
with $j=i-1$ for which the
third of conditions \eqref{vD} also holds. In this case,
\begin{eqnarray}\nn 0&=&D^T_{\alpha\beta}v_\alpha^{(i)}v_\beta^{(i-1)}
=-\sum_{k\ne i}^{N}D^T_{\alpha\beta}v_\alpha^{(k)}v_\beta^{(i-1)}\\
\nn &=&
-D^T_{\alpha\beta} v_\alpha^{(i-2)}v_\beta^{(i-1)}-D^T_{\alpha\beta}
v_\alpha^{(i-1)}v_\beta^{(i-1)}\\
&=&-D^T_{\alpha\beta}v_\alpha^{(i-2)}
v_\beta^{(i-1)}-v^{(i-1)}\cdot v^{(i-1)}\end{eqnarray}
and
 \begin{equation}
0=D^T_{\alpha\beta}v_\alpha^{(i-1)}v_\beta^{(i-2)}=D_{\alpha\beta}
v_\alpha^{(i-2)}v_\beta^{(i-1)}.\end{equation}
From these (on resetting $i\rightarrow i+1$) one may deduce
\begin{equation}
2v^{(i-1)}\cdot v^{(i)}+v^{(i)}\cdot v^{(i)}=0,\end{equation}
and hence \begin{equation}
E_{\alpha\beta}v_\alpha^{(i-1)}v_\beta^{(i)}=v^{(i-1)}\cdot v^{(i)}.\end{equation}
At this stage, all the inner products of the vectors (up to an
overall scale factor) have been deduced. Taking $N=r+1$, it is clear the inner products
are those between simple roots of $A_r$, and thus it is possible to put
 $v^{(i)}\sim\alpha_i$. The action of the matrix $E$ on any pair of vectors
  $v^{(i)},$ $v^{(j)},$ is known and, after a little algebra, it is found that in terms
of the fundamental weights $\lambda_{i}$ of $A_{r}$
\begin{equation}
E=\sum_{i=1}^{r}(\lambda_{i+1}\lambda_{i}^{T}-\lambda_{i}\lambda_{i+1}^{T}),\end{equation}
where $\lambda_{r+1}=\lambda_{N}=0.$ Also, it is found that \begin{equation}
D=1-E=2\sum_{i=1}^{r}\lambda_{i}(\lambda_{i}^{T}-\lambda_{i+1}^{T}),\end{equation}
as before. At this point,  everything is now essentially known and
the bulk affine Toda potentials will be recovered from \eqref{BVW}.

\section{One soliton solution}
To investigate what happens to a single soliton solution the
defect conditions \eqref{affineconditions} for the affine
$a^{(1)}_{r}$ Toda theories will be used in the context of complex
valued fields. Using Hirota's formalism \cite{Hiro80},
\cite{Hollowood92}
 the single soliton solution can be written in the following manner
\begin{equation}
\phi_{(a)}=-\sum^{r}_{j=0}\alpha_{j}\ln \tau_{j}, \;\;\;
\tau_{j}=1+E_a\omega^{aj}, \;\;\;\;
\psi_{(a)}=-\sum^{r}_{j=0}\alpha_{j}\ln \sigma_{j}, \;\;\;
\sigma_{j}=1+\Lambda E_a\omega^{aj},
\end{equation}
with
$$E_a=e^{a_a x+b_a t +\xi}, \;\;\;\;\; (a_a,\ b_a)=m_{a}
(\cosh{\theta},\sinh{\theta}), \;\;\;\;\; m_{a}^2=
a_a^{2}-b_a^{2}= 4\sin^{2}\left(\frac{a\pi}{r+1}\right),$$
and $\omega=\exp(2\pi i/(r+1))$.
It is presumed the rapidity $\theta$ and the type of soliton $a$
is the same on both sides of the defect and the additional factor $\Lambda$
represents the
effect of the defect on the soliton as it passes through. Using
the properties of D \eqref{dperm}, and multiplying on the left by
$\alpha_{k}$, the following two equations which the single
soliton solution has to satisfy are obtained,
\begin{eqnarray*}
&&
2\frac{\tau^{'}_{k}}{\tau_{k}}-\frac{\tau^{'}_{k+1}}{\tau_{k+1}}-
\frac{\tau^{'}_{k-1}}{\tau_{k-1}}+\frac{\dot{\tau}_{k+1}}{\tau_{k+1}}-
\frac{\dot{\tau}_{k-1}}{\tau_{k-1}}-2\frac{\dot{\sigma}_{k}}{\sigma_{k}}
+2\frac{\dot{\sigma}_{k-1}}{\sigma_{k-1}} \\
&&
-d\left(\frac{\tau_{k+1}}{\tau_{k}}\frac{\sigma_{k-1}}{\sigma_{k}}
-\frac{\tau_{k}}{\tau_{k-1}}\frac{\sigma_{k-2}}{\sigma_{k-1}}\right)-
\frac{1}{d}\left(\frac{\tau_{k-1}}{\tau_{k}}\frac{\sigma_{k}}
{\sigma_{k-1}}-\frac{\tau_{k}}{\tau_{k+1}}\frac{\sigma_{k+1}}
{\sigma_{k}} \right)=0, \\
&&
2\frac{\sigma^{'}_{k}}{\sigma_{k}}-\frac{\sigma^{'}_{k+1}}{\sigma_{k+1}}-
\frac{\sigma^{'}_{k-1}}{\sigma_{k-1}}+\frac{\dot{\sigma}_{k+1}}{\sigma_{k+1}}-
\frac{\dot{\sigma}_{k-1}}{\sigma_{k-1}}-2\frac{\dot{\tau}_{k}}{\tau_{k}}
+2\frac{\dot{\tau}_{k+1}}{\sigma_{k+1}} \\
&&
+d\left(\frac{\tau_{k+1}}{\tau_{k}}\frac{\sigma_{k-1}}{\sigma_{k}}
-\frac{\tau_{k+2}}{\tau_{k+1}}\frac{\sigma_{k}}{\sigma_{k+1}}\right)-
\frac{1}{d}\left(\frac{\tau_{k-1}}{\tau_{k}}\frac{\sigma_{k}}
{\sigma_{k-1}}-\frac{\tau_{k}}{\tau_{k+1}}\frac{\sigma_{k+1}}
{\sigma_{k}} \right)=0.
\end{eqnarray*}
Substituting the expressions for $\tau$ and $\sigma$ reveals two
equations in $E$ of order five, which are compatible, and satisfied by
the following expression for $\Lambda$
\begin{equation}
\Lambda=\left(\frac{ie^{\theta}+de^{i\zeta_a}}{ie^{\theta}+de^{-i\zeta_a}}\right),
\;\;\;\;\;\;\;\; \zeta_a=\frac{\pi a}{r+1}.
\end{equation}
Thus, the effect of the defect is to delay or advance the soliton
as it passes. For the case of a self-conjugate soliton,
that is for a soliton corresponding to $a=(r+1)/2$ (with $r$ odd),  the
delay is real and identical with the expression already found for the sine-Gordon
soliton \cite{bczlandau}$$\Lambda=\left(\frac{e^{\theta}+d}{e^{\theta}-d}\right).$$
On the other hand, for a conjugate pair of solitons, $a$ and $\bar{a}=
(r+1-a)$, the values
of the delays are given by
\begin{equation}
\Lambda_{(a)}=\left(\frac{ie^{\theta}+de^{i\zeta_a}}{ie^{\theta}+de^{-i\zeta_a}}\right),
\;\;\;\;\;\;\;\;
\Lambda_{(\bar{a})}=\left(\frac{ie^{\theta}-de^{-i\zeta_a}}{ie^{\theta}-
de^{i\zeta_a}}\right)\equiv \bar{\Lambda}_a.
\end{equation}
As already noted in \cite{bczlandau},  self-conjugate solitons can be absorbed or
emitted for special values of real rapidity for which the delay or its
inverse is infinite.

\p
Taking the
other boundary conditions, that is those that allow the spin -2 charge
to be conserved instead, the corresponding expression for $\Lambda$
is obtained by replacing $\theta$ with $-\theta$.

\section{Summary and discussion}

The purpose of this article has been to explore further the ideas introduced in
\cite{bczlandau} and analyse the extent to which the Lagrangian description of a defect
applies to affine Toda field theories other than the sine/sinh-Gordon model. Curiously,
the $a_r^{(1)} $ models appear to be singled out, which is not quite what was anticipated
since the affine Toda field theories using any of the affine root data have
hitherto had rather similar properties with relatively small differences of detail.
Perhaps a lack of sufficient imagination is the real problem and only time will tell.
On the other hand, the description of a defect is very close to the B\"acklund
transformation idea, and  B\"acklund transformations were found first
for the $a_r^{(1)} $ models \cite{Fordy80},
being rather more subtle for the other affine Toda field
theories. After first noticing
the association between a defect and a B\"acklund transformation it was expected its
characteristically striking property of creating or annihilating solitons might show up
in the guise of permitting a defect to change soliton number (for the
use of B\"acklund transformations in this context for the $a_r^{(1)}$ case,
see \cite{Liao93}). After all, a
genuine defect will violate topological charge since there is no requirement for
a field (or its space derivative) to be continuous through a defect.
However, that interesting possibility does not appear to be realised.

\p
Another intriguing possibility is the potential for controlling solitons by suitably
tuning delays. It is known there are physical situations where the sine-Gordon
solitons are relevant (a Josephson junction, for example) and it is interesting
to speculate on what kinds of physically realisable mechanisms
might produce a defect of the type described here. If such mechanisms exist then
they could be used to control the arrival times of signals propagated by solitons.
There is already literature concerning the effects of impurities on the motion of
`solitons' within various models (see, for example, \cite{Alexeeva99}) although
the emphasis is rather different from that of the approach adopted in this article.

\p
It was not the purpose of this article to explore the associated quantum field theories.
However, there are two interesting domains. The real field theories, whose particle
spectrum is a set of scalar particles, distinguished by the eigenvalues of the higher
spin charges, which are likely to have a set of transmission phase factors $T_a(\theta_a)$, one
for each particle type, dependent upon the defect parameter and on the bulk coupling
constant in addition to the rapidity of the particle. It is expected that as the
bulk coupling tends to zero the transmission factor approaches its classical
value dependent upon the defect parameter and rapidity. One interesting question is
the fate of the weak-strong coupling duality under the transformation of the bulk
coupling $\beta \rightarrow 4\pi /\beta$. On the other hand,
the complex field theory,
whose particle spectrum is conjectured to be a set of multiplets of sizes coinciding with the
dimensions of the fundamental representations of $A_r$ \cite{Hollowood92}
(although not all of these
have corresponding classical `soliton' solutions \cite{McGhee93}),
is expected to have matrix transmission
factors, which will also satisfy a Yang-Baxter type relation to ensure compatibility with
the bulk S-matrix on either side of the defect, in addition to having a
suitable classical limit. Determining these various quantum properties is a matter
for further investigations beyond the scope of the present article.

\vskip 1cm \noindent{\bf Acknowledgements} \vskip .5cm \noindent
One of us (CZ) is supported by a University
of York Studentship. Another (EC) wishes to thank the Asia Pacific Center for
Theoretical Physics for its hospitality during the closing stages
of preparing this article. The work has been performed under the auspices of
EUCLID - a European Commission funded TMR Network - contract
number HPRN-CT-2002-00325. We have all benefited from occasional
discussions with Gustav Delius and Evgeny Sklyanin.


\end{document}